\documentclass[11pt]{article}
\usepackage{geometry}
\usepackage{amsmath,amssymb}
\usepackage{amsfonts}
\usepackage{amsthm}
\usepackage{indentfirst}

\usepackage{bm}
\usepackage{graphicx}
\usepackage{ascmac} 
\usepackage{multicol}	
\usepackage{enumerate}
\usepackage{wrapfig}
\usepackage{subfigure}
\usepackage{amsbsy}
\usepackage{dcolumn}
\newcolumntype{I}{!{\vrule width 3\arrayrulewidth}}
\newlength\savedwidth
\newcommand\whline{%
    \noalign{\xdef\origarrayrulewidth{\the\arrayrulewidth}%
    \global\arrayrulewidth 3\arrayrulewidth}%
    \hline%
    \noalign{\global\arrayrulewidth\origarrayrulewidth}%
}

\numberwithin{equation}{section}

\theoremstyle{plain}

\theoremstyle{definition}

\theoremstyle{remark}

\begin{document}
\thispagestyle{empty}


\begin{center}
{\sc\large Takashi Arima}, \
{\sc\large Tommaso Ruggeri},  \ {\small and}  \
{\sc\large Masaru Sugiyama}
\end{center}
\vspace {1.1cm}


\begin{center}
    \large{\textbf{Dispersion Relations of Longitudinal and Transverse Waves \\ in a Rarefied Polyatomic Gas \\ based on Rational Extended Thermodynamics}}
\end{center}

\renewcommand{\thefootnote}{\fnsymbol{footnote}}


\renewcommand{\thefootnote}{\arabic{footnote}}
\setcounter{footnote}{0}

\vspace{0.6cm}
\begin{center}
\begin{minipage}[t]{11cm}
\small{
\noindent \textbf{Abstract.}
We present a complete analysis of the dispersion relations for longitudinal and transverse waves in a rarefied polyatomic gas based on Rational Extended Thermodynamics (RET), which describes the evolution of a non-polytropic gas in nonequilibrium. Observability of the second mode of the longitudinal wave and the transverse wave is discussed although these waves are usually not payed much attention. The cases of CO$_2$ gas and para-H$_2$ gas are specifically analyzed as typical examples.
\medskip

\noindent \textbf{Keywords.}
dispersion relation, rarefied polyatomic gas, longitudinal wave, transverse wave, rational extended thermodynamics, hyperbolic system of balance laws
\medskip

\noindent \textbf{Mathematics~Subject~Classification:}
76N30, 76P05, 35L60.

}
\end{minipage}
\end{center}

\bigskip


\section{Introduction}

Rational Extended Thermodynamics (RET) is a theory that aims to describe highly nonequilibrium phenomena in viscous and heat-conducting gases in terms of a hyperbolic system of field equations. It was originally motivated by the moment equations associated with the distribution function in the Boltzmann equation for rarefied monatomic gases.
In order to obtain the closed system of field equations, universal principles such as the entropy principle, the Galilean invariance, and the convexity of entropy density are systematically utilized. At the molecular level, the closure procedure corresponds to the use of the Maximum Entropy Principle.  
RET of rarefied monatomic gases is now a well-established theory, and many works are summarized in the book by M\"uller and Ruggeri  \cite{RET}.

After the publication of the book \cite{RET}, Arima, Taniguchi, Ruggeri, and Sugiyama \cite{ETdense} established a RET theory of rarefied polyatomic gases, where two hierarchies of field equations were introduced. The recent studies of RET for both classical and relativistic polyatomic gases are reviewed in the books by Ruggeri and Sugiyama \cite{RS1,RS2}.


Theoretical predictions derived from RET concerning various nonequilibrium phenomena in a gas such as shock wave, and light scattering have been successfully confirmed by experimental data. For more details, see \cite{RS2} and references therein. The study of the dispersion relation of longitudinal sound wave \cite{dispersion}, in particular, has remarkably clarified the applicability range of RET. In the limit of short relaxation time, we obtain the Navier-Stokes-Fourier (NSF) theory from the RET theory. In this sense, the applicability range of RET is wider than that of NSF.



The aim of the present paper is to derive the dispersion relations of longitudinal and transverse waves propagating in three-dimensional space. The latter wave was excluded in the previous study \cite{dispersion} in which we considered only one-space dimension. We also analyze the second mode of the longitudinal wave. Although it is usually said that the second mode of the longitudinal wave and the transverse wave are not observed in experiments, we show that the dissipation effects, which result from the relaxation of the fields, provide a possibility of observing these waves.


\section{RET theory with 14 fields}

In this section, we summarize briefly the RET theory with 14 fields (RET$_{14}$) of rarefied polyatomic gases.

\subsection{Thermal and caloric equations of state}

The thermal and caloric equations of state of a rarefied polyatomic gas are given by
\begin{align*}
& p = \frac{k_B}{m}\rho T \ \ \ \mathrm{and} \ \ \ \varepsilon \equiv \varepsilon(T),  
\end{align*}
where $p$, $\rho$, $T$ and $\varepsilon$ are pressure, mass density, absolute temperature, and 
specific internal energy, respectively. $k_B$ and $m$ are the Boltzmann constant and the mass of a molecule.  Note that gases are, in general, non-polytropic, that is, the specific heat 
at constant volume, 
\begin{align*}
 &c_v  = \frac{ \mathrm{d} \varepsilon}{ \mathrm{d} T},
\end{align*}
has not a constant value but depends on the temperature.

\subsection{Linearized basic equations of RET$_{14}$}

We assume that a nonequilibrium state of a gas is prescribed by the 14 independent field variables 
${\boldsymbol{u}} \equiv (\rho, v_i, T, \Pi, \sigma_{\langle ij\rangle}, q_i$), where $v_i$, $\Pi (= -\sigma_{ii}/3)$, $\sigma_{\langle ij\rangle}$, and $q_i$ are, respectively, velocity, dynamic (nonequilibrium) pressure, symmetric traceless part of the viscous stress $\sigma_{ij}$, and heat flux \cite{ETdense}.

Since we are now focusing on a wave with small amplitude, we linearize the closed system of field equations of RET$_{14}$ around an equilibrium state: ${\boldsymbol{u}}_0 \equiv (\rho_0,0,T_0,0,0,0)$. Then the linearized system for the perturbed field $ {\bar{\boldsymbol{u}}} (={\boldsymbol{u}}-{\boldsymbol{u}}_0)$ is given by (we omit the bar from now on) 
\begin{align}
 \begin{split}
  &\frac{\partial \rho}{\partial t } + \rho_0\frac{\partial  {v}_k}{\partial x_k} = 0,\\[3pt]
  &\rho_0 \frac{\partial  {v}_i}{\partial t} + \frac{k_B}{m}T_0 \frac{\partial  {\rho}}{\partial x_i}+\frac{k_B}{m}\rho_0 \frac{\partial  {T}}{\partial x_i}+\frac{\partial  {\Pi}}{\partial x_i}
  - \frac{\partial  {\sigma}_{\langle ij\rangle}}{\partial x_j}  = 0,\\[3pt]
  &\frac{k_B}{m}\rho_0 {\hat{c}_v} \frac{\partial  T}{\partial t}
  + \frac{k_B}{m}\rho_0 T_0 \frac{\partial  {v}_k}{\partial x_k}
  + \frac{\partial  {q}_k}{\partial x_k}=0, \\[3pt]
  &\frac{\partial  \Pi}{\partial t} +  \frac{2\hat{c}_v -3}{3\hat{c}_v}\frac{k_B}{m}\rho_0 T_0\frac{\partial  {v}_k}{\partial x_k}
  + \frac{2\hat{c}_v-3 }{3 \hat{c}_v(1+\hat{c}_v)}\frac{\partial  {q}_k}{\partial x_k}
  = - \frac{1}{\tau_{\Pi}} {\Pi},\\[3pt]
  &\frac{\partial  {\sigma}_{\langle ij\rangle}}{\partial t}
   - 2\frac{k_B}{m}\rho_0 T_0 \frac{\partial  {v}_{\langle i}}{\partial x_{j\rangle}}
   - \frac{2}{1+\displaystyle{\hat{c}_v}} \frac{\partial  {q}_{\langle i}}{\partial x_{j\rangle}} = - \frac{1}{\tau_{\sigma}} {\sigma}_{\langle ij\rangle} , \\[3pt]
  & \frac{\partial  {q}_i}{\partial t} 
  + \left(1+\displaystyle{\hat{c}_v}\right) \left(\frac{k_B}{m}\right)^2 \rho_0 T_0  \frac{\partial  {T}}{\partial x_i}
  + \frac{k_B}{m}T_0 \frac{\partial  {\Pi}}{\partial x_i}
  -\frac{k_B}{m}T_0 \frac{\partial  {\sigma}_{\langle ik\rangle}}{\partial x_k}
  = -\frac{1}{\tau_{q}}  {q}_i  , 
 \end{split}\label{systemET}
\end{align}
 where $\hat{c}_v $ is the dimensionless specific heat at the reference equilibrium state: 
\begin{align*}
 &\hat{c}_v  = \left. \frac{c_{v}}{k_B/m} \right|_{T=T_0} .
\end{align*}
The system \eqref{systemET} is in the form of a linear hyperbolic system of the type:
\begin{equation}\label{sistema}
    \frac{\partial   {\boldsymbol{u}}}{\partial t} + {\boldsymbol{A}}_0^i \, \frac{\partial  {\boldsymbol{u}}}{\partial x_i} = {\boldsymbol{B}}_0  {\boldsymbol{u}}.
\end{equation}
The relaxation times $\tau_{\Pi}, \tau_{\sigma}$, and $\tau_{q}$ in Eq. \eqref{systemET}, which are functions of $\rho$ and $T$, are also evaluated at the reference equilibrium state.

By conducting the Maxwellian iteration \cite{RET,ETdense,Ikenberry}, we obtain the relations between the relaxation times and the bulk viscosity $\nu$, shear viscosity $\mu$, and heat conductivity $\kappa$ as follows:
\begin{align*}
 &\nu = \frac{2\hat{c}_v -3}{3\hat{c}_v}\frac{k_B}{m}\rho_0 T_0 \tau_{\Pi} ,
 \ \ \  \mu = \frac{k_B}{m}\rho_0 T_0 \tau_{\sigma} , \ \ \  \kappa = \left(1+\displaystyle{\hat{c}_v}\right)\left(\frac{k_B}{m}\right)^2 \rho_0 T_0 \tau_{q} . 
\end{align*}


\section{Dispersion relations}

In this section, we derive the dispersion relations and then obtain the high-frequency limit of the phase velocity and the attenuation factor.

\subsection{Dispersion relation, phase velocity, and attenuation factor}

We now study a plane harmonic wave propagating in a direction ${\boldsymbol{n}}\equiv (n_i)$ in three-dimensional space with the frequency $\omega$ 
and the complex wave number $k =\Re(k ) + \mathrm{i} \Im(k ) $ such that 
\begin{align}
 &  {\boldsymbol{u}}=\delta\boldsymbol{u} \, \, \mathrm{e}^{\mathrm{i} (\omega t - k n_i x_i)},   \label{solve}
\end{align}
where $\delta{\boldsymbol{u}}$ is a constant amplitude vector. 
Substituting \eqref{solve} into   \eqref{sistema} we obtain the following algebraic linear system:
%
 \begin{align}
 & \left(-\omega \boldsymbol{I} + k \boldsymbol{A}_{0n} - \mathrm{i} \boldsymbol{B}_0\right)\delta \, {\boldsymbol{u}} = {\bf{0}}, \qquad ({\boldsymbol{A}}_{0n} = {\boldsymbol{A}}_0^i n_i).\label{linear_eq}
 \end{align}
Explicit form of the algebraic system \eqref{linear_eq} can be obtained by applying the following formal substitution rule to the system \eqref{sistema}:
\begin{equation*}
    \frac{\partial}{\partial t } \rightarrow -\omega \delta, \qquad \frac{\partial}{\partial x_i } \rightarrow k n_i  \delta, \qquad \boldsymbol{B}_0 \boldsymbol{u} \rightarrow \mathrm{i} \boldsymbol{B}_0 \delta \boldsymbol{u}.
\end{equation*}
 Then we have
  \begin{align}\label{mioalg}
 & -\omega \delta \rho + k \rho_0 \delta v_n =0, \nonumber \\
 & -\omega \delta v_i + \frac{k}{\rho_0}\left\{n_i\left(\frac{k_B }{m}T_0\delta \rho+ \frac{k_B }{m}\rho_0\delta T + \delta \Pi\right) -  \delta \sigma_{ni}\right\} =0,\nonumber \\
  & -\omega \delta T +  \frac{k}{\hat{c}_v}\left(T_0 \delta v_n + \frac{1}{\frac{k_B }{m} \rho_0}\delta q_n\right) = 0,\nonumber \\
   & -\omega \delta \Pi +  k \frac{2\hat{c}_v -3  }{3 \hat{c}_v}\left(\frac{k_B }{m} \rho_0 T_0\delta v_n + \frac{1}{1+\hat{c}_v}\delta q_n\right) = -\frac{\mathrm{i}}{\tau_\Pi} \delta \Pi,\\
    & -\omega \delta {\sigma}_{\langle ij\rangle}- k \left\{\frac{k_B }{m}\rho_0 T_0\left(n_j \delta v_i+n_i \delta v_j -\frac{2}{3} \delta v_n \delta_{ij}\right)\right. +
    \nonumber \\
    & \hspace{2.5cm} \left. + \frac{1}{1+\hat{c}_v}\left(n_j \delta q_i+n_i \delta q_j -\frac{2}{3} \delta q_n \delta_{ij}\right)
   \right\} = -\frac{\mathrm{i}}{\tau_\sigma} \delta {\sigma}_{\langle ij\rangle}, \nonumber\\
   & -\omega \delta q_i + k \left\{(1+\hat{c}_v) \left(\frac{k_B }{m}\right)^2 \rho_0 T_0 n_i \delta T + \frac{k_B}{m}T_0\left(n_i \delta \Pi - \delta \sigma_{ni}\right)\right\} = -\frac{\mathrm{i}}{\tau_q} \delta q_i, \nonumber
 \end{align}
 where  
 \begin{equation*}
     \delta v_n = \delta v_i n_i, \quad \delta q_n = \delta q_i n_i, \quad \delta \sigma_{ni} = \delta \sigma_{\langle ij\rangle}n_j.
 \end{equation*}
 Multiplying \eqref{mioalg}$_{2,6}$ by $n_i$ and \eqref{mioalg}$_{5}$ by $n_i n_j$, we obtain
 the homogeneous algebraic system:
 \begin{equation*}
 \left(-\omega \boldsymbol{I}^{(L)} + k\boldsymbol{A}_{0n}^{(L)} - {\mathrm{i}}\boldsymbol{B}_0^{(L)} \right)\delta  \boldsymbol{u}^{(L)}  ={\bf{0}}
 \end{equation*}
with $\boldsymbol{I}^{(L)}$ being $6\times 6$ identity matrix, and 
\begin{align*} 
 &\boldsymbol{A}_{0n}^{(L)} =
\left(
\begin{array}{cccccc}
  0 & \rho_0  & 0 & 0 & 0 & 0 \\
 \displaystyle{\frac{k_B}{m}\frac{ T_0}{\rho_0 }} & 0 &\displaystyle \frac{k_B}{m} & \displaystyle{\frac{1}{\rho_0 }} & -\displaystyle{\frac{1}{\rho_0} } & 0  \\
 0 & \displaystyle{\frac{T_0}{{\hat{c}_v}}} & 0 & 0 & 0 & \displaystyle{\frac{1}{\frac{k_B}{m} {\hat{c}_v} \rho_0 }}  \\[12pt]
 0 &  \displaystyle{\frac{2\hat{c}_v -3}{3\hat{c}_v} \frac{k_B}{m} \rho_0 T_0}  & 0 & 0 & 0 & \displaystyle{\frac{2\hat{c}_v-3}{3 \hat{c}_v(1+\hat{c}_v)}}  \\[12pt]
 0&  \displaystyle{-\frac{4}{3} \frac{k_B}{m} \rho_0 T_0} & 0 & 0 & 0 &  -\displaystyle{\frac{4}{3(1+\displaystyle{\hat{c}_v})}}  \\
 0 & 0 & \displaystyle{\left(1+{\hat{c}_v}\right)\left(\frac{k_B}{m}\right)^2 \rho_0 T_0}  & \ \   \displaystyle\frac{k_B}{m} T_0 \ \ & \ \  \displaystyle - \frac{k_B}{m}T_0 \ \  & 0  
\end{array}
\right),
\end{align*}
\begin{align*} 
 &\boldsymbol{B}_0^{(L)} = \left(
\begin{array}{cccccc}
 0 & 0 & 0 & 0 & 0 & 0 \\
 0 & 0 & 0 & 0 & 0 & 0  \\
 0 & 0 & 0 & 0 & 0 & 0 \\
 0 & 0 & 0 & \displaystyle{-\frac{1}{\tau _{\Pi}}} & 0 & 0   \\
 0 & 0 & 0 & 0 & \displaystyle{-\frac{1}{\tau _{\sigma}}} & 0  \\
 0 & 0 & 0 & 0 & 0 & \displaystyle{-\frac{1}{\tau_{q}}} \\
\end{array}
\right), \quad \delta \boldsymbol{u}^{(L)} \equiv (\delta \rho, \delta v_n, \delta T, \delta \Pi, \delta \sigma_{nn}, \delta q_n )^T,
\end{align*}
where
\begin{align*}
    \delta \sigma_{nn} = \delta \sigma_{\langle ij\rangle}n_i n_j.
\end{align*}
Then we have the dispersion relation:
\begin{align}
 &\mathrm{det} \left( - \omega \boldsymbol{I}^{(L)} + k \boldsymbol{A}_{0n}^{(L)} - {\mathrm{i}}\boldsymbol{B}_0^{(L)} \right) =0. \label{DRL}
\end{align}

If \eqref{DRL} is not zero, then $\delta \boldsymbol{u}^{(L)} \equiv (\delta \rho, \delta v_n, \delta T, \delta \Pi, \delta \sigma_{nn}, \delta q_n )^T={\bf{0}}$ and \eqref{mioalg} reduces to
 \begin{align*}
 \begin{split}
 & -\omega \delta v_i -  \frac{k}{\rho_0} \delta \sigma_{ni}  =0, \\
    & -\omega \delta {\sigma}_{ni}- k \left( \frac{k_B}{m}\rho_0 T_0 \delta v_i  +
   \frac{1}{1+\hat{c}_v}  \delta q_i \right)
    = -\frac{\mathrm{i}}{\tau_\sigma} \delta {\sigma}_{ni},\\
   & -\omega \delta q_i - k   \frac{k_B}{m} T_0 \delta \sigma_{ni}   = -\frac{\mathrm{i}}{\tau_q} \delta q_i,
   \end{split}
 \end{align*}
or, in a compact form, 
 \begin{equation*}
  \left(-\omega \boldsymbol{I}^{(T)} + k \boldsymbol{A}_{0n}^{(T)} - {\mathrm{i}}\boldsymbol{B}_0^{(T)} \right) \delta \boldsymbol{u}^{(T)} = {\bf{0}},  
 \end{equation*}
where $\boldsymbol{I}^{(T)}$ is $3\times 3$ identity matrix and 
\begin{align*} 
 &\boldsymbol{A}_{0n}^{(T)} =
\left(
\begin{array}{ccc}
0 &\displaystyle-\frac{1}{\rho_0} & 0\\
 \displaystyle - \frac{k_B}{m}\rho_0 T_0 & 0  & \displaystyle- \frac{1}{1+\hat{c}_v} \\
 0& \displaystyle - \frac{k_B}{m}T_0& 0
\end{array}
\right),  \qquad  \boldsymbol{B}_0 ^{(T)} = \left(
\begin{array}{ccc}
  0  &  0 & 0 \\
 0 & \displaystyle{-\frac{1}{\tau_{\sigma}}}  & 0 \\
0 &  0 & \displaystyle{-\frac{1}{\tau_{q}}}  
\end{array}
\right), \\
&\delta \boldsymbol{u}^{(T)} \equiv \left(\delta v_i, \delta \sigma_{ni}, \delta q_i \right)^T.  
\end{align*}
This is the system for the transverse wave with the corresponding  dispersion relation:
\begin{align}
\mathrm{det} \left( - \omega \boldsymbol{I}^{(T)} + k \boldsymbol{A}_0^{(T)} - {\mathrm{i}}\boldsymbol{B}_0^{(T)} \right) = 0.  \label{DRT}
\end{align}

\bigskip

The phase velocity $v_{ph}$ and the attenuation factor $\alpha$ are 
defined as
\begin{align*}
 &v_{ph}(\omega)=\frac{\omega}{\Re (k)}
,       \qquad
 \alpha(\omega)=-\Im (k). 
\end{align*}
In addition, it is useful to introduce the attenuation per wavelength $\alpha_{\lambda}$:
\begin{align*}
 & \alpha_{\lambda}(\omega)= \alpha \lambda=\frac{2\pi v_{ph} \alpha }{\omega} = -2\pi \frac{{\cal I}m (k)}{{\cal R}e (k)},
\end{align*}
where $\lambda$ is the wavelength. 

In agreement with the general result  \cite{RuggeriMuracchini1992}, the phase velocity and the attenuation factor in the high-frequency limit are given by
\begin{align}
 &v_{ph}^{(\infty)} \equiv \lim_{\omega \rightarrow \infty} v_{ph}(\omega) = U_0, \label{limit_v}\\
 &\alpha^{(\infty)} U_0 \equiv \lim_{\omega \rightarrow \infty} \alpha(\omega) U_0 = - \boldsymbol{l}_0 
 \cdot \boldsymbol{B}_0 \cdot \boldsymbol{d}_0,    \label{limit_a}
\end{align}
where  $U_0$ is an  eigenvalue of $\boldsymbol{A}_{0n} $ (characteristic velocity),
and $\boldsymbol{l}_0$ and $\boldsymbol{d}_0$ are the corresponding left and right eigenvectors.

\subsection{Longitudinal wave}
By introducing the dimensionless parameters defined by
\begin{align*}
 &\Omega = \tau_{\sigma} \omega,  \ \ \ \hat{\tau}_q=\frac{\tau_{q}}{\tau_{\sigma}}\left(=\left(1+{\hat{c}_v}\right)^{-1}\frac{m\kappa}{k_B \mu}\right) , \ \ \
 \hat{\tau}_\Pi= \frac{\tau_{\Pi}}{\tau_{\sigma}}\left(= \frac{3\hat{c}_v}{2\hat{c}_v-3}\frac{\nu}{\mu}\right), 
\end{align*}
the dispersion relation \eqref{DRL} is shown explicitly as 
\begin{align}
 &\frac{{\hat{c}_v}(c_0 z)^4}{3 \Omega ^2 \left(1+{\hat{c}_v}\right)^2 \hat{\tau}_\Pi}
 \left\{
 -3 (1+{\hat{c}_v})-\mathrm{i} \Omega 
 \left(3+7{\hat{c}_v}+5{\hat{c}_v}\hat{\tau}_\Pi\right)+9 \Omega ^2 {\hat{c}_v} \hat{\tau}_\Pi\right\}  \nonumber \\
 &+ \frac{(c_0 z)^2}{3 \Omega ^3 (1+{\hat{c}_v})^2 \hat{\tau}_q \hat{\tau}_\Pi}
 \left\{-3\mathrm{i}\left(1+{\hat{c}_v}\right)^2 +
  \right. \nonumber\\
 &\ \quad +\Omega \left(1+{\hat{c}_v}\right) 
  \left[3+7{\hat{c}_v}+5{\hat{c}_v}\hat{\tau}_\Pi +6\left(1+{\hat{c}_v}\right)\hat{\tau}_q \right] + \nonumber\\
 &\ \quad  +\mathrm{i} \Omega^2 \left[2\left(3+10{\hat{c}_v}+5{\hat{c}_v}^2\right)\hat{\tau}_q + 9{\hat{c}_v}\left(1+{\hat{c}_v}\right)\right.\hat{\tau}_\Pi
 + \left. {\hat{c}_v}\left(13+8{\hat{c}_v}\right)\hat{\tau}_q\hat{\tau}_\Pi\right] +
 \label{dispersion_ip}  \\
 &\ \quad      -3\Omega^3 {\hat{c}_v}\left(7+4{\hat{c}_v}\right)\hat{\tau}_\Pi\hat{\tau}_q\Big\}  \nonumber \\
 &+\frac{(\Omega -\mathrm{i}) (\hat{\tau}_\Pi\Omega -\mathrm{i}) (\hat{\tau}_q \Omega -\mathrm{i})}{ \Omega ^3\hat{\tau}_\Pi\hat{\tau}_q} = 0 \nonumber 
\end{align}
with $z=k/\omega$  and $c_0$ being the sound velocity in an equilibrium state defined by
\begin{align*}
 &c_0 
 =\sqrt{\frac{\hat{c}_v + 1}{\hat{c}_v}} c_s,
\end{align*}
where
\begin{align*}
  c_s = \sqrt{\frac{k_B}{m}T_0} .
\end{align*}
Therefore, for given $\hat{c}_v$, $\hat{\tau}_q$, and $\hat{\tau}_\Pi$, the quantity $c_0 z (=c_0 k/\omega)$ is calculated 
from Eq. \eqref{dispersion_ip} as the function of $\Omega$.

From \eqref{limit_v} and \eqref{limit_a}, we have the high-frequency limit of the phase velocity and the attenuation factor as follows:
\begin{align*}
 v_{ph}^{(L),(\infty)} =& \pm \sqrt{\frac{  4{\hat{c}_v} + 7 + F}{2(1+\hat{c}_v)}}c_s,  \\ 
 \alpha^{(L),(\infty)} = & \pm
 \frac{\sqrt{2 (1+{\hat{c}_v})^{3}} \left\{ F \left(4+{\hat{c}_v}\right)- 22 - 11 {\hat{c}_v} + 2 {\hat{c}_v}^2\right\}}
 {9 \hat{c}_v \tau_{\sigma} c_s   \sqrt{7+4 {\hat{c}_v} + F} \left(7 + 4 {\hat{c}_v} - F\right)^2 F} \\
 &\quad  \left\{4{\hat{c}_v}+ \frac{3 {\hat{c}_v}\left(8+2 {\hat{c}_v}-F\right)}{\hat{\tau}_q} +\frac{-3+2{\hat{c}_v}}{\hat{\tau}_\Pi} \right\} ,  
\end{align*}
where
\begin{align*}
 &F= \pm \sqrt{37+ 32 {\hat{c}_v} +4{\hat{c}_v}^2} .
\end{align*}
When $F$ is positive (negative), the wave represents the first (second) mode. 

\begin{figure}[ht!]
\centering
   \includegraphics[width=60mm]{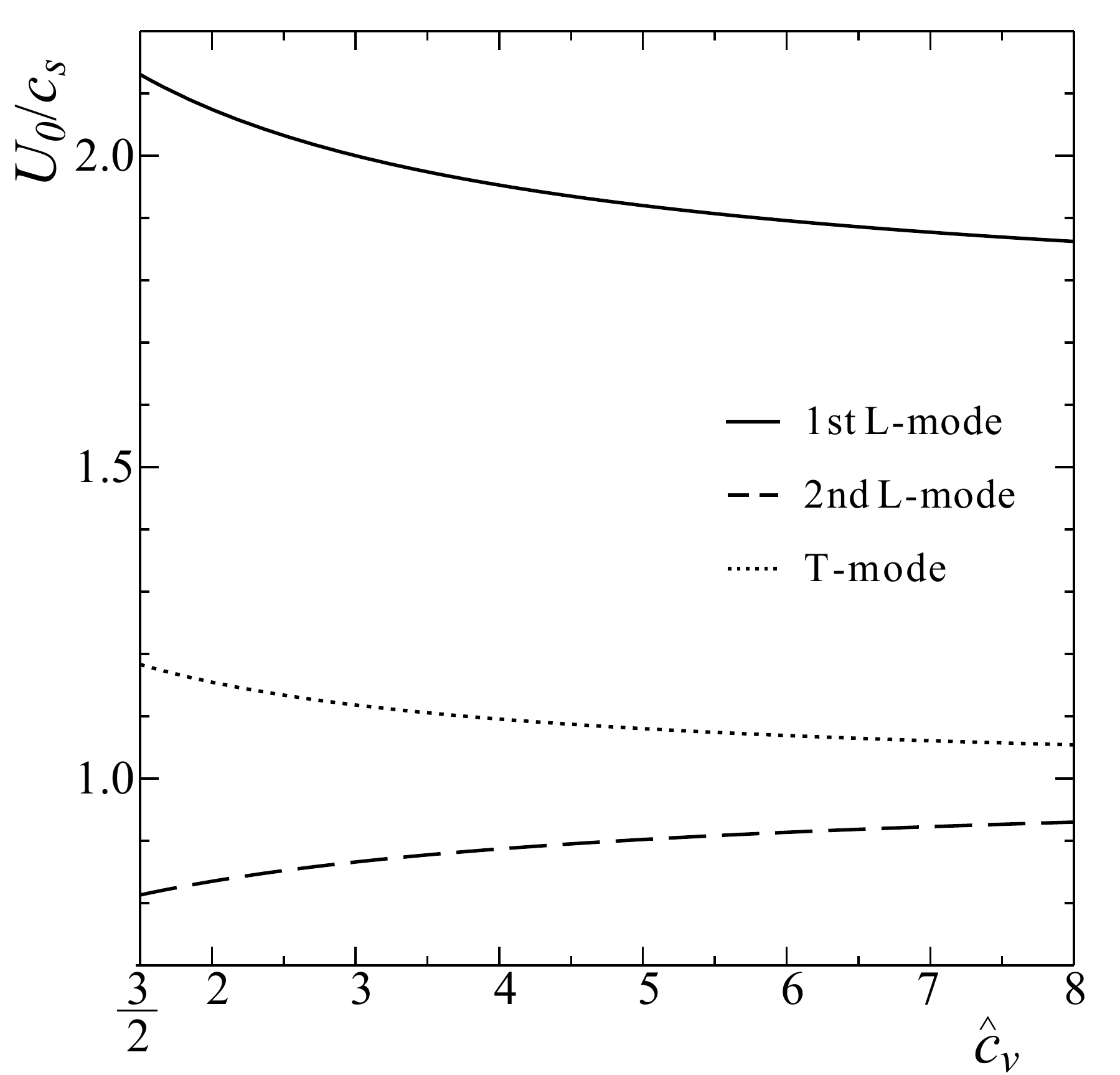}
\caption{Dependence of $v_{ph}^{(\infty)}/c_s (=U_0/c_s)$ on the specific heat for the first longitudinal mode (solid line), the second longitudinal mode (dashed line), and the transverse mode (dotted line).}
  \label{fig:U0}
\end{figure}

We notice that $v_{ph}^{(L),(\infty)}/c_s (=U_0/c_s)$ depends only on $\hat{c}_v$. Its dependence is shown in Fig. \ref{fig:U0} both for the first and second modes. 
In particular, in a monatomic gas with $\hat{c}_v=3/2$, $v_{ph}^{(L),(\infty)}$ of the first and second modes are, respectively, $2.13051c_s$ and $0.812975c_s$. In the limit where $\hat{c}_v \rightarrow \infty$, $v_{ph}^{(L),(\infty)}$ of the first and second modes approach $\sqrt{3}c_s$ and $c_s$, respectively. 
On the other hand, $\alpha^{(L),(\infty)}$ depends not only on $\hat{c}_v$ but also on the relaxation times.  
In a monatomic gas, $\alpha_\lambda^{(L),(\infty)}$ of the first and second modes are, respectively, $(0.0951852+0.0931368/\hat{\tau}_q )/({\tau}_\sigma c_s)$ and $( 0.0238989+0.370948 / \hat{\tau}_q)/({\tau}_\sigma c_s)$.
In the limit of large specific heat, $\alpha^{(L),(\infty)}$ approaches $\frac{1+2\hat{\tau}_\Pi}{9\sqrt{3}\hat{\tau}_\Pi\tau_\sigma c_s}$ and $\frac{1}{2 \hat{\tau}_q \tau_\sigma c_s}$, respecively.

\subsection{Transverse wave}

From \eqref{DRT}, we obtain
\begin{align*}
 &\left(c_s z\right)^2 \left(\frac{\hat{c}_v+2}{\hat{c}_v+1}-\frac{\mathrm{i}}{\hat{\tau}_q \Omega }\right)-\frac{(\Omega -\mathrm{i}) (\hat{\tau}_q \Omega -\mathrm{i})}{\hat{\tau}_q \Omega ^2}=0.
\end{align*}

From \eqref{limit_v} and \eqref{limit_a}, we have the high-frequency limit of the phase velocity and the attenuation factor as follows:
\begin{align*}
 &v_{ph}^{(T),(\infty)} = \pm \sqrt{\frac{\hat{c}_v+2}{\hat{c}_v+1}}c_s, \\
 &\alpha^{(T),(\infty)} = \pm \frac{\sqrt{\hat{c}_v+1}\left\{1+(\hat{c}_v+2) \hat{\tau}_q\right\}}{2 (\hat{c}_v+2)^{3/2} \hat{\tau}_q \tau_\sigma c_s}.
\end{align*}

Similar to the case of longitudinal waves, the dependence of $v_{ph}^{(T),(\infty)}/c_s$ on the specific heat is shown in Fig. \ref{fig:U0}. $v_{ph}^{(T),(\infty)}$ is $\sqrt{7/5}c_s$ in a monatomic gas and approaches $c_s$ in the limit of $\hat{c}_v \rightarrow \infty$ which is the same as that of the second mode.
On the other hand, $\alpha^{(T),(\infty)}$ is $(0.422577 +0.120736 /\hat{\tau}_q )/({\tau}_\sigma c_s)$ in a monatomic gas and approaches $\frac{1}{2 \tau_\sigma c_s}$  in the limit of $\hat{c}_v \rightarrow \infty$.
\bigskip

{\bf Remark} - 
The phase velocities in the limit of high frequency coincide with the characteristic eigenvalues of the system \eqref{sistema} (see \cite{RuggeriMuracchini1992}), and, according to the fact that the system is symmetric hyperbolic at least in the neighborhood of equilibrium, all the eigenvalues are real. While, in the case far from equilibrium as is usual with RET, the hyperbolicity remains valid only in a domain of nonequilibrium variables called {\em  hyperbolicity region} (see for more details \cite{RS2}). For the present model of non-polytropic gas, this region was determined very recently in \cite{Brini2023}.

\section{Examples: CO$_2$ gas and para-H$_2$ gas}

\begin{figure}[ht!]
\centering
\begin{minipage}[t]{0.44\textwidth}
   \includegraphics[width=0.9\textwidth]{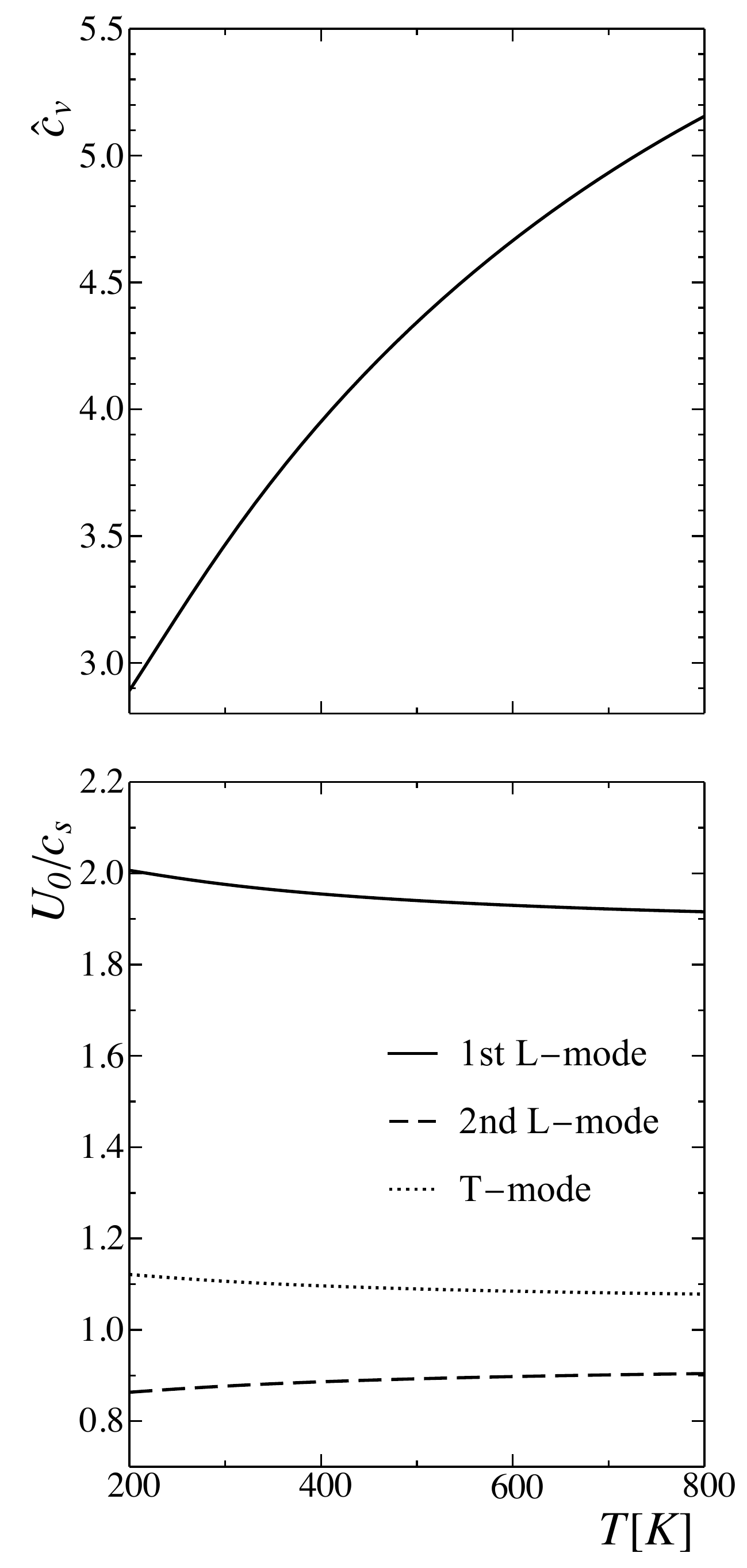}
\end{minipage}
\ 
\begin{minipage}[t]{0.44\textwidth}
   \includegraphics[width=0.9\textwidth]{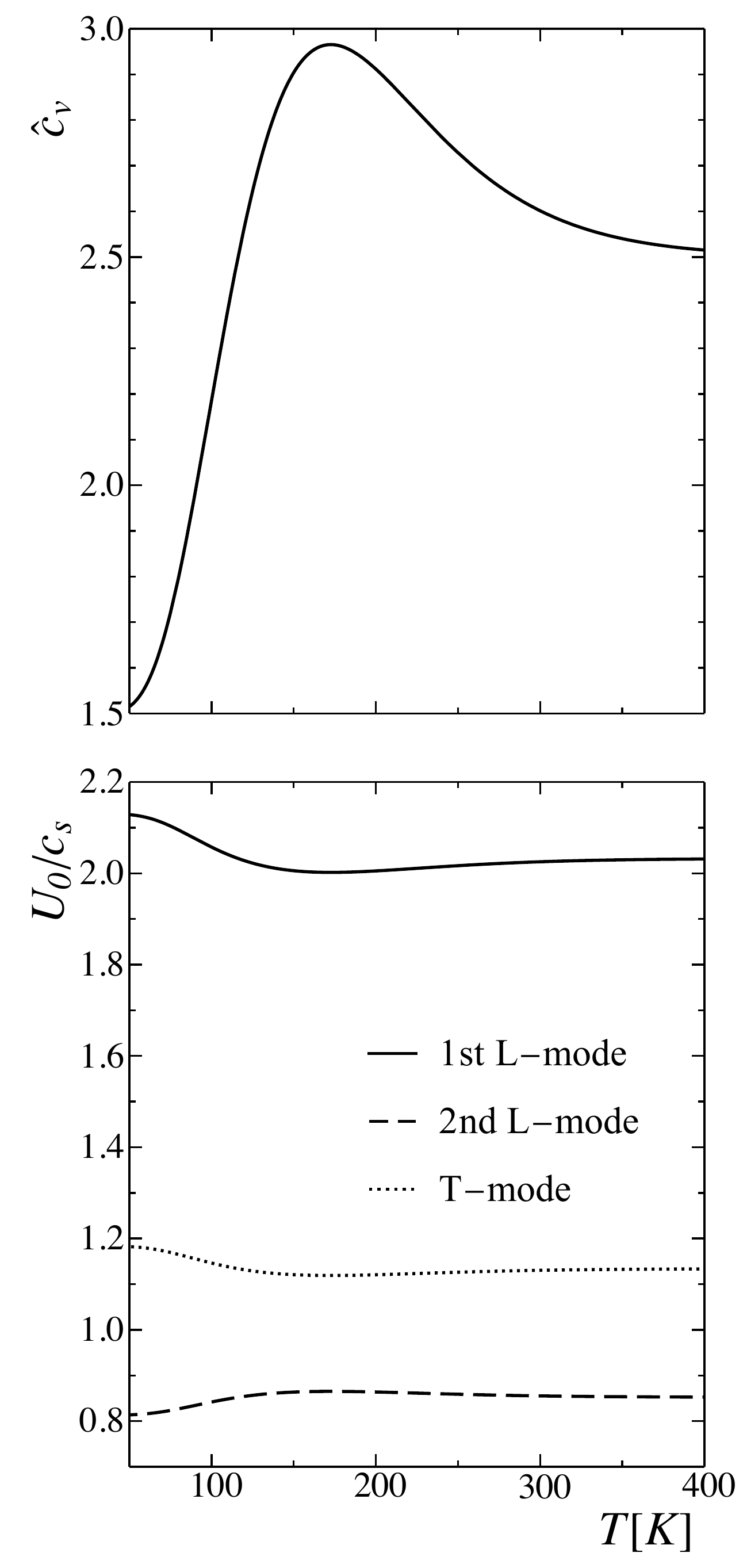}
\end{minipage}
\caption{Temperature dependence of the specific heat (up) and $U_0/c_s$ (down) for CO$_2$ gas (left) and para-H$_2$ gas (right).  In the figure of $U_0/c_s$, the solid, dashed, and dotted lines correspond, respectively, to the first longitudinal mode, the second longitudinal mode, and the transverse mode.}  
\label{fig:U-H2}    
\end{figure}

As typical examples, we study the waves in CO$_2$ and para-H$_2$ gases. The temperature dependence of the specific heat is determined on the basis of statistical mechanics \cite{LandauSM,LandauQM} as shown in Fig.\ref{fig:U-H2}. The temperature ranges in the figure are characterized as follows: the molecular vibrational modes are excited for CO$_2$ gas and the molecular rotational mode is excited for para-H$_2$ gas. The temperature dependence of $U_0/c_s$ for both gases is also shown in Fig.\ref{fig:U-H2}.

\begin{figure}[ht!]
\centering
\begin{minipage}[t]{0.49\textwidth}
\centering
   \includegraphics[width=\textwidth]{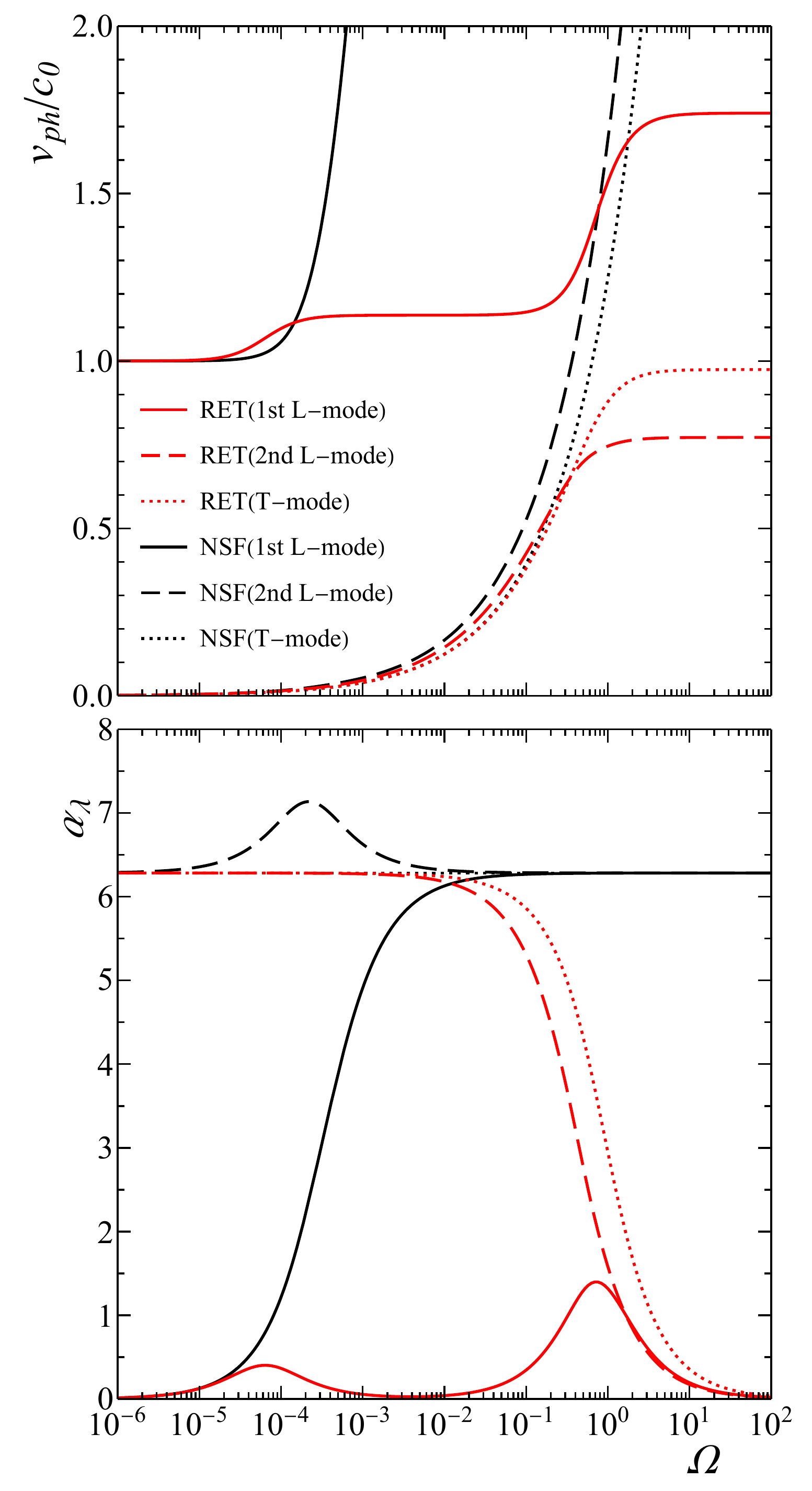}
\end{minipage}
\hfill
\begin{minipage}[t]{0.49\textwidth}
\centering
   \includegraphics[width=\textwidth]{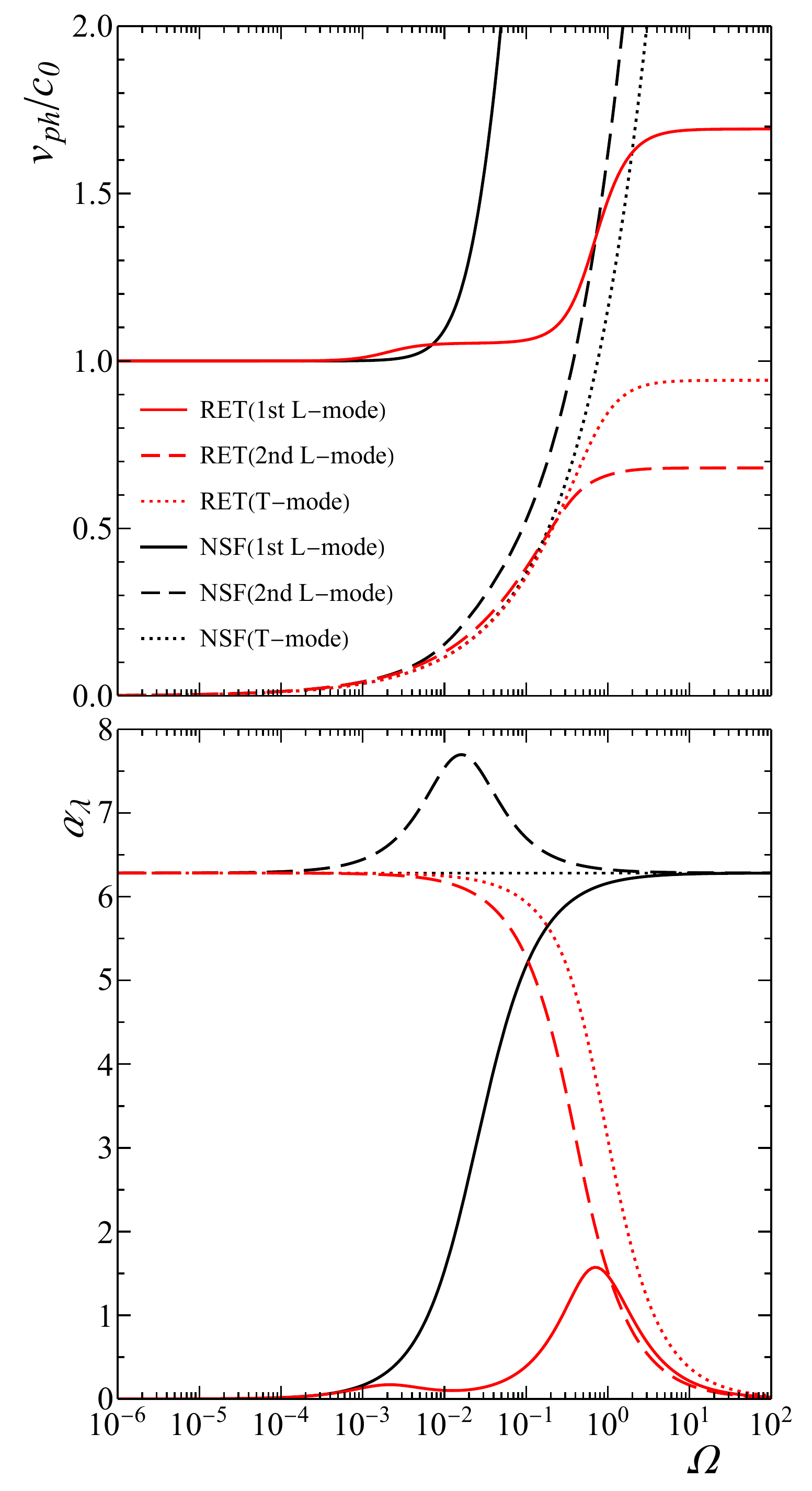}
\end{minipage}
\caption{Dependence of the normalized phase velocity $v_{ph}/c_0$ (up) and attenuation per wavelength $\alpha_\lambda$ (down) on the dimensionless frequency $\Omega$ for CO$_2$ gas at $T_0=293$K (left) and para-H$_2$ gas at $T_0=90.2$K (right) predicted by RET$_{14}$ (red) and  NSF (black). The solid, dashed, and dotted lines correspond, respectively, to the first longitudinal mode, the second longitudinal mode, and the transversal mode.}
  \label{fig:wave}
\end{figure}

Fig.\ref{fig:wave} shows the frequency dependence of the phase velocity and the attenuation per wavelength predicted by the RET$_{14}$ theory and the NSF theory.  For CO$_2$ gas, we have adopted the reference equilibrium state at $T_0=295$K and $p_0=69$mm Hg. Then, we have $\hat{c}_v = 3.45$ and the relaxation times estimated as $\tau_\Pi = 2.2 \times 10^{-5}$ [s], $\tau_\sigma = 1.6 \times 10^{-9}$ [s], and $\tau_q = 2.2 \times 10^{-9}$ [s] \cite{ET14shock}. For para-H$_2$ gas, we have taken the reference equilibrium state at $T_0=90.2$K. In this case, we have $\hat{c}_v = 1.99$ and the relaxation times estimated as $\tau_\Pi p_0 = 1.85 \times 10^{-3}$ [s$\cdot$ Pa], $\tau_\sigma p_0 =3.97 \times 10^{-6}$ [s$\cdot$ Pa], and $\tau_q p_0 = 5.16 \times 10^{-6}$ [s$\cdot$ Pa] \cite{dispersion}. We remark that, in these gases, the relaxation time of the dynamic pressure is several orders of magnitude larger than the other relaxation times.  
We find that despite the different molecular excitation processes in CO$_2$ and para-H$_2$ gases, the frequency dependence of $v_{ph}$ and $\alpha_\lambda$ is similar to each other and the difference of the values between the two gases is determined by the specific heat and the ratio of the relaxation times.

In the prediction of the first longitudinal mode by RET for both gases, due to the dissipation process of the dynamic pressure with large value of $\tau_\Pi$, we find a steep change of $v_{ph}$ and a peak of $\alpha_\lambda$ around $\Omega \sim \hat{\tau}_\Pi^{-1}$. Around $\Omega \sim \hat{\tau}_q^{-1} \sim 10^0$, we find another steep change of $v_{ph}$ and peak of $\alpha_\lambda$. On the other hand, we notice that the prediction by NSF is completely different. We already discussed the different predictions by the theories in detail in \cite{dispersion} and shown that the prediction of RET is consistent with experimental data.

Regarding the second longitudinal mode, RET$_{14}$ predicts that, different from the first mode, near equilibrium, $v_{ph} \sim 0$ and $\alpha_\lambda \sim 2\pi$.  As the frequency increases ($\Omega \gtrsim 10^{-2}$), these values vary monotonically. While the predictions by RET and NSF theories agree with each other in the small frequency region, they diverge at high frequencies.  The transverse mode predicted by RET and NSF exhibits similar behavior. 

We comment on the potential observability of the second longitudinal mode and the transversal mode through their frequency-dependence of $\alpha_\lambda$. At low frequencies, $\alpha_\lambda$ is large, which indicates the wave is absorbed over a short distance. Conversely, for high frequencies ($\Omega \gtrsim 10^0$), values of $\alpha_\lambda$ of these two modes approach the value of the first longitudinal mode. This suggests that these modes may be observable by high-frequency experiments.
 Only  RET enables such theoretical predictions, as the predictions of $\alpha_\lambda$ by NSF exceed $2\pi$ for any frequency.



\section{Conclusion}

In this paper, we have studied the dispersion relations for the first and second modes of the longitudinal wave and the transverse wave using the model of RET with $14$ fields for a polyatomic non-polytropic gas. Contrary to the usual assertion that both the second mode of the longitudinal wave and the transverse wave are not observed in experiments, we have pointed out 
the possibility of observing these waves in the high-frequency region.

\vspace{0.8cm} \indent {\it
A\,c\,k\,n\,o\,w\,l\,e\,d\,g\,m\,e\,n\,t\,s.\;} 
 This paper is dedicated to the memory of our colleague and unforgettable friend Giampiero Spiga. The work has been partially supported by JSPS KAKENHI Grant Numbers 22K03912 (T.A.).  
The work has also been carried out in the framework of the activities of the Italian National Group of Mathematical Physics of the Italian National Institute of High Mathematics GNFM/INdAM (T.R.).


\bigskip
\begin{center}


\end{center}

\bigskip
\bigskip
\begin{minipage}[t]{10cm}
\begin{flushleft}
\small{
\textsc{Takashi Arima}
\\*National Institute of Technology, 
\\*Tomakomai College, 
\\*443 Nishikioka
\\*Tomakomai 059-1275, Japan
\\*e-mail: arima@tomakomai-ct.ac.jp
\\[0.4cm]
\textsc{Tommaso Ruggeri}
\\*Department of Mathematics,
\\*University of Bologna,
\\*Via Saragozza, 8 
\\*40139 Bologna,  Italy
\\*e-mail: tommaso.ruggeri@unibo.it
\\[0.4cm]
\textsc{Masaru Sugiyama}
\\*Nagoya Institute of Technology, 
\\*Gokiso-cho, Showa-ku
\\* Nagoya 466-8555, Japan
\\*e-mail: sugiyama@nitech.ac.jp
}
\end{flushleft}
\end{minipage}


\end{document}